\documentstyle[epsfig,palatino,mathpple,latexsym]{mn}

\def\etal{{et~al.\/ }}

\def\ltsima{$\; \buildrel < \over \sim \;$}
\def\simlt{\lower.5ex\hbox{\ltsima}}
\def\gtsima{$\; \buildrel > \over \sim \;$}
\def\simgt{\lower.5ex\hbox{\gtsima}}
\def\simless{\mathbin{\lower 3pt\hbox
   {$\rlap{\raise 5pt\hbox{$\char'074$}}\mathchar"7218$}}}   % < or of order
\def\simgreat{\mathbin{\lower 3pt\hbox
   {$\rlap{\raise 5pt\hbox{$\char'076$}}\mathchar"7218$}}}   % > or of order
\def\ib{$I_{814}$\/ }
\def\vb{$V_{606}$\/ }
\def\bb{$B_{450}$\/ }
\def\ub{$U_{300}$\/ }
\def\iw{$I_{814W}$\/ }

\def\vi{$V_{606}-I_{814}$\/ }
\def\bv{$B_{450}-V_{606}$\/ }
\def\ub{$U_{300}-B_{450}$\/ }

\def\dvi{$\delta(V-I)$\/ }
\def\sfrdensity{$\rho_{\mathrm SFR}$\/ }

\def\apj{ApJ\/ }
\def\mnras{MNRAS\/ }

\title[Evolving spheroidals in the Hubble Deep Fields]{Evidence for
Evolving Spheroidals in the Hubble Deep Fields North and South}

\author[F. Menanteau, R. G. Abraham \& R. S. Ellis]{
        F. Menanteau $^1$\thanks{fmenante@ast.cam.ac.uk},
        R. G. Abraham$^{1,2}$ and
        R. S. Ellis$^{1,3}$\\
$^1$Institute of Astronomy, University of Cambridge, Madingley Road,
Cambridge CB3 0HA, England\\
$^2$Astronomy Department, University of Toronto, 60 St. George Street, Toronto, ON, M5S 3H8, Canada.\\ 
$^3$Astronomy Department, Caltech, 105-24, Pasadena, CA 91125, USA.}
\date{Received:\ \ \ Accepted: }
\pagerange{\pageref{firstpage}--\pageref{lastpage}}
\pubyear{2000}
\begin{document}

\maketitle

\label{firstpage}

\begin{abstract}

We investigate the dispersion in the {\em internal} colours of faint
spheroidals in the Hubble Deep Fields North and South. In high
redshift rapid-collapse scenarios, the dispersion in internal colours
should be small at moderate redshift apart from a small
metallicity-induced reddening in the enriched cores. However,
recently-assembled spheroidals are likely to show non-homologous
internal colours, at least until younger stellar populations become
fully mixed. Here we find that a remarkably large fraction ($\simgt
30$\%) of the morphologically-classified spheroidals with
$I_{814W}<24$~mag show strong variations in internal colour, which we
take as evidence for recent episodes of star-formation. In most cases
these colour variations manifest themselves via the presence of blue
cores, an effect of opposite sign to that expected from metallicity
gradients.  Examining similarly-selected ellipticals in five rich
clusters with $0.37<z<0.83$ we find a significant lower dispersion in
their internal colours. This suggests that the colour inhomogeneities
have a strong environmental dependence being weakest in dense
environments where spheroidal formation was presumably accelerated at
early times. We use the trends defined by the cluster sample to define
an empirical model based on a high-redshift of formation and estimate
that at $z \sim 1$ about half the field spheroidals must be undergoing
recent episodes of star-formation. Using spectral synthesis models, we
construct the time dependence of the density of star-formation
($\rho_{\mathrm SFR}$).  Although the samples are currently small, we
find evidence for an increase in \sfrdensity between $z=0$ to
$z=1$. We discuss the implications of this rise in the context of that
observed in the similar rise in the abundance of galaxies with
irregular morphology. Regardless of whether there is a connection our
results provide strong evidence for the continued formation of field
spheroidals over 0$<z<$1.

\end{abstract}


\section{INTRODUCTION}

The age distribution of elliptical galaxies remains a central topic in
galaxy evolution.  For many years ellipticals were viewed as old
systems which formed as the result of a short and intense burst of
star formation (Baade 1957, Sandage 1986), after which the stars
within them passively evolved.  Strong observational support for this
view came from old and coeval ellipticals found in rich clusters at
low (Bower Lucey \& Ellis 1992) and high redshifts (Ellis \etal 1997,
Stanford \etal 1997).

However, in hierarchical models of galaxy formation dominated by cold
dark matter (CDM), elliptical galaxies form over a longer period, as
the result of mergers between low mass disks (Kauffmann \etal 1996,
Baugh \etal 1996). In these models clusters form from the highest peaks
of the density fluctuations, so the existence of old, coeval cluster
ellipticals is not necessarily in contradiction with hierarchical
models. Clearly, regions of high density are poorly suited for testing
models for elliptical galaxy formation.

The controversy regarding the epoch of formation of elliptical
galaxies has prompted several observational campaigns designed to
establish whether field ellipticals share the same evolutionary
history as their clustered counterparts.  Recent tests
have focused on the very red optical-IR colours predicted by 
single-collapse models. Zepf (1997) and Barger \etal (1998) adopted a
colour-based approach by searching for a very red tail in the
optical-IR colour distribution of faint HDF galaxies.  Very few sources
were found with colours matching those expected for high redshifts
of formation. Similarly, Menanteau \etal (1999), using a
statistically-complete sample of morphologically-selected
spheroidals from Hubble Space Telescope (HST) archival data, concluded
on the basis of optical-infrared colours that field ellipticals cannot 
have formed the bulk of their stars in a
single-burst of formation at a high redshift. 

However, an early period of collapse may be rescued if massive systems
subsequently suffer minor episodes of star formation rendering the
observed colours bluer than the limits explored in the above
observational studies. Jim\'enez \etal (1998) propose a multi-zone
model of spheroidal formation which predicts bluer colours over the
redshift range explored. This is consistent with recent studies by
Abraham \etal (1999) and Kodama, Bower \& Bell (1998, hereafter KBB)
which found a large fraction ($\sim30\%$) of spheroidals have
properties which differ from those predicted by single-collapse
high-redshift models. More recently Tamura \etal (2000) studied the
origin of colour gradients in bright elliptical galaxies in the
Northern HDF where they report a similar fraction ($\sim30\%$) of
systems with blue integrated colours. However, they claim that colour
gradients are very small and originated by stellar metallicity.

The ability of HST to resolve distant galaxies, and the very deep
multi-colour photometry available in the Hubble Deep Fields (HDFs),
opens up new possibilities for probing the internal characteristics of
distant galaxies. In this paper we adopt a methodology similar to that
used in Abraham \etal (1999), where the spatially-resolved colours of
galaxies of known redshift in the Northern HDF were used to probe
their star-formation history.  We improve on Abraham \etal (1999) by
comparing the internal properties of field spheroidal with their
clustered counterparts and by enlarging the sample of early type
galaxies by a factor of 6, including the incorporation of high
signal-to-noise (S/N) data from the Southern HDF.

A plan for the paper follows.  In $\S$\ref{sample-sec}
we present the field and cluster spheroidal sample. In
$\S$\ref{model-free-sec} we introduce the principles of our
methodology and discuss our results and their implications for the
evolution of spheroidals under a free-model approach. In
$\S$\ref{model-sec} we attempt to model the observed colour
dispersions, estimate the corresponding time-scales for the
star-formation activity and compute the present SFR for spheroidals
from the HDFs. In $\S$\ref{conclusions} we summarize our conclusions.

\section{FIELD AND CLUSTER SAMPLES}
\label{sample-sec}

\subsection{Spheroidals in the Hubble Deep Fields}
\label{field-sample-sec}

The Hubble Deep Fields (HDFs) provide our primary sample of field
faint spheroidals. In the present paper sources in the HDF fields are
designated using the IAU IDs given in version 2 of the HDF {\tt
SExtractor} catalogues constructed by the Space Telescope Science
Institute. Vega magnitudes are used throughout this paper.  HDF
galaxies were morphologically classified using both the automated
methodology described in Menanteau \etal (1999) (based on both, the
central concentration ($C$) and asymmetry ($A$) parameters from
Abraham \etal 1996), as well as using the visual classifications made
by one of us (RSE). Visual and automated classification have been
shown to agree well in previous studies (eg. Brinchmann \etal 1998;
Menanteau \etal 1999). However, Marleau and Simard (1998) claim that a
significant fraction of galaxies visually classified as ellipticals in
the van der Bergh \etal (1996) morphological catalog of the HDF were
disk-dominated galaxies, although most of these systems are fainter
than our selection limits.

A limiting magnitude of \ib = 24 mag was adopted, so that information
for each galaxy is distributed over a large enough number of pixels to
allow us to define meaningful measures of our model-independent
estimator that we will describe in section $\S$\ref{model-free-sec}.  We augment
the spectroscopic redshifts in our sample using photometric redshift
estimates kindly provided by S.  Gwyn (Gwyn \etal 2000 in
preparation).  Photometric redshifts play a particularly important
role in the Southern HDF field, where the published spectroscopic data
is presently rather limited. Our final sample consists of 79
spheroidal galaxies, 24 of which have spectroscopic redshifts and 55
of which have photometric redshifts.

\subsection{Cluster Spheroidals}

\label{cluster-sample-sec}

\begin{table}
\centering
\caption{The HST Cluster Sample}
\begin{tabular}{@{}llll}

\hline
Cluster     & Filter 1 & Filter 2 & $z$ \\
\hline\hline
A370            &       F814W   &       F555W   & 0.37 \\
Cl0412-65       &       F814W   &       F555W   & 0.51 \\
Cl0016+16       &       F814W   &       F555W   & 0.55 \\
Cl0054-27       &       F814W   &       F555W   & 0.56 \\
MS1054+03       &       F814W   &       F606W   & 0.83 \\
\hline
\end{tabular}
\label{cluster_table}
\end{table}

In order to compare field and cluster samples directly, we selected a
sample of high-redshift cluster ellipticals observed in two passbands
with HST's Wide Field Planetary Camera 2 (WFPC2) over a similar
wavelength baseline to \vi and spanning a redshift range comparable to
the one observed for field spheroidals. From the MORPHS collaboration
dataset \cite{Smail97}, we selected 4 rich cluster at
$0.37<z<0.56$. We also used the data from van Dokkum (1999) for the
X-ray cluster MS1054+03 at $z=0.83$.

Both studies have published morphological classifications, from which
we select spheroidals as objects classified as E and E/S0 (but not
S0).  In Table~\ref{cluster_table} we list the properties of the
selected clusters. It is important to note that the cluster imaging
data is not as deep as the HDF data, and to allow for this our cluster
sample is restricted to be brighter than a limiting magnitude of
$I_{814}=22$~mag, ie. two magnitudes brighter than the corresponding
HDF field data.  We discuss the potential limitations introduced by
adopting different cluster and field limiting magnitudes further below.
Another difference is that the MORPHS sample was imaged using F555W
instead of the F606W filter which defines the $V$-band in the HDF. We
defer a detailed study of its influence on our analysis until the
next section.

\section{MODEL-INDEPENDENT ANALYSIS}
\label{model-free-sec}

\subsection{Methodology}
\label{methodology}

In some ways our methodology for examining the internal colours of an
individual galaxy can be considered as a generalization of the use of
colour-magnitude relations in studying the history of ellipticals in
galaxy clusters: the photometric {\em dispersion} is a powerful probe
of variations in star-formation history provided systematic effects
are under control (Bower, Lucey \& Ellis 1992).

\begin{figure}
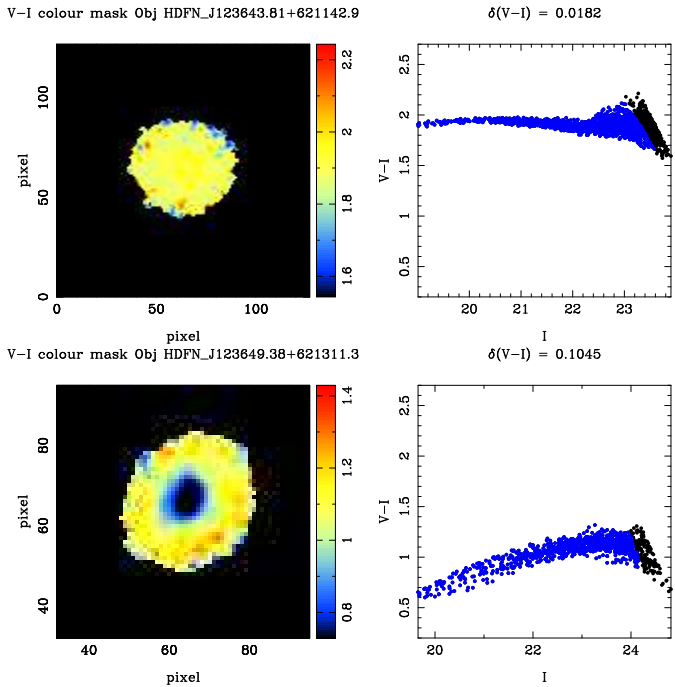

\begin{center}
\leavevmode
\centerline{\psfig{file=oldE.ps,width=4.5cm,angle=-90}}
\centerline{\psfig{file=blueE.ps,width=4.5cm,angle=-90}}
\end{center}
\caption{\em The methodology of internal colour variations as
applied to two HDF-N spheroidals. The upper panel shows an
$I_{814}=20.48$~mag example with low internal scatter at $z=0.77$, 
whereas the bottom panel illustrates an $I_{814}=21.66$ example  
with a bluer core at $z=0.48$. The pixel-by-pixel colour distributions
are shown alongside each case. Coloured dots refer to pixels whose
SNR is above the adopted threshold.}
\label{examples}
\end{figure}

Initially we will focus on a largely model-independent approach to the
problem of understanding how the internal colours of field ellipticals
vary with redshift relative to those of their clustered counterparts.
We will concentrate only on \ib and \vb~bands, as the resulting \vi
colours have substantially smaller observational errors than the \bv
and \ub colours, especially for systems dominated by old stellar
populations.

\subsubsection{The Homogeneity Estimator}
\label{estimator}
\begin{figure}
\begin{center}
\leavevmode
\centerline{\psfig{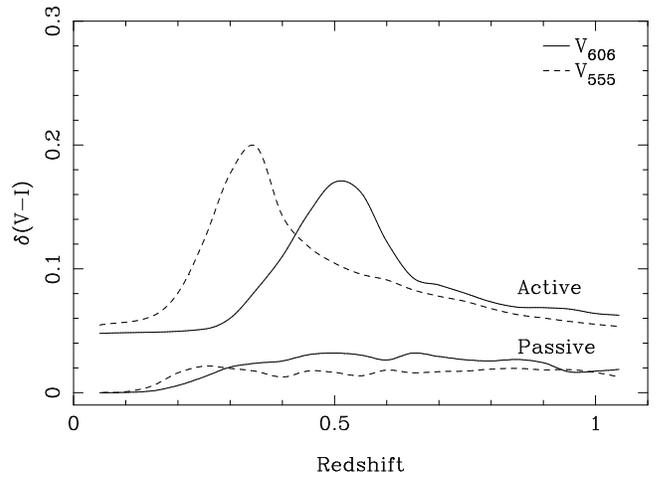}}
\end{center}
\caption{\em The observed \dvi as a function of redshift for modelled
galaxies viewed at various redshifts with two photometric colour
systems. The \vi (solid lines) and \mbox{$V_{555}-I_{814}$} (dashed
line) curves are plotted for two cases: a single-burst (``Passive'')
system corresponding to an epoch of formation $z_F=5$, and an
``Active'' case which incorporates a 15\% by mass burst of star
formation centrally super-imposed on an old galaxy at the time of
observation.}
\label{simfilters-fig}
\end{figure}

After registering the Northern HDF images to sub-pixel accuracy, we
isolated each galaxy from the background sky by selecting all
contiguous pixels within a surface brightness threshold of
\mbox{$V_{606}=25.5$~mag~arcsec$^{-2}$}. We select this limit in order
to maximize the signal/noise per pixel associated to \vi colours.

The distribution of \vi~colours for individual pixels in the galaxy is
used to calculate a quantity $\delta(V-I)$ which characterises
the internal homogeneity of a galaxy. A $\delta(V-I)$
statistic is defined as:
\begin{equation}
\delta(V-I) = 2N\frac{\sum ( x_{i} - \bar{x})^{2}S(x_i){\mathit{SNR}}(x_i)}{\sum S(x_i){\mathit{SNR(x_i)}}}
\end{equation}
\begin{equation}
\bar{x} = \frac{\sum x_{i}S(x_i){\mathit{SNR}}(x_i)}{\sum S(x_i){\mathit{SNR}}(x_i)}
\label{dvi-eqn}
\end{equation}
where $S(x)$ is a selection function for pixels with signal/noise
above a certain threshold ($SNR>1.3$) such that $S(x)=1$ for pixels
above and $S(x)=0$ for pixel below the threshold. $SNR(x_i)$ is the
signal-to-noise ratio for a given pixel colour $x_i$, and $N$ is an
arbitrary scale factor. The selection function $S(x)$ and the
weighting according to $SNR(x)$ address biases arising from noise
variations at pixel scales by rejecting low signal pixels and
weighting the pixels contribution proportionally to their
signal. Figure~\ref{examples} show an example of spheroidals in HDF-N
with both high and low \dvi.

As mentioned previously, the MORPHS dataset was observed with a
different filter baseline than that used for the HDF. To quantify the
effect of this difference on $\delta(V-I)$, we used stellar population
synthesis libraries (Bruzual \& Charlot 1996, BC96) to calculate the
observed colours (in both photometric systems) for two galaxy types:
{\em a)} an old single-burst elliptical (formation epoch $z_F=5$ and
$e$-folding time $\tau=1$~Gyr) and, {\em b)} a star-forming
elliptical galaxy consisting of a recent burst ($100$~Myr) centrally
superimposed on an old system. In order to emulate the geometrical
properties of the modelled galaxies, we use the observed colours at a
given redshift to create elliptical galaxies with $r^{1/4}$ profiles
using a customised version of the {\sc IRAF} package {\sc artdata} and
a fixed physical size with a half-light radius $r_e=2.5$~kpc. For the
case of the blue core, we attempt to match the properties of
Figure~\ref{examples} lower panel, with a burst that accounts for 15\%
of the total mass of the galaxy with and a physical size of $0.1r_e$.
\dvi for both photometric systems are shown as a function of redshift
in Figure~\ref{simfilters-fig}.  This exercise demonstrates that the
substitution of F555W for F606W does not have a major impact on the
calculation of \dvi. The slightly longer wavelength baseline in the
cluster \dvi estimates should result in a greater sensitivity to
recent activity.

It is important to note that because \dvi is based on {\em observed}
(rather than rest-frame) colours, its sensitivity to star formation is
redshift-dependent. Unfortunately, it is not possible to use additional
filters to create a rest-frame equivalent since the shorter wavelength
HDF data has poorer signal/noise and the cluster data only consists of
two passbands. In order to quantify this restriction, we studied values
of \dvi obtained for the blue core case modelled in
Figure~\ref{simfilters-fig}.  We placed this modelled galaxy, at the
epoch corresponding to its maximum inhomogeneity (representing the most
extreme case we will discuss) at various redshifts calculating \dvi at
each redshift.  Figure~\ref{simfilters-fig} shows increased sensitivity
to the fixed star formation rate at $z\sim0.5$ for F606W and $z\sim0.4$
for F555W; otherwise the sensitivity is broadly constant. This increase
arises as the 4000\AA~ break in the underlying passive system passes
through the F606W and F555W filters and will not seriously compromise
our analysis.

\subsubsection{The Influence of the HST Point-Spread Function}
\label{psf-section}

\begin{figure}
\begin{center}
\leavevmode
\centerline{\psfig{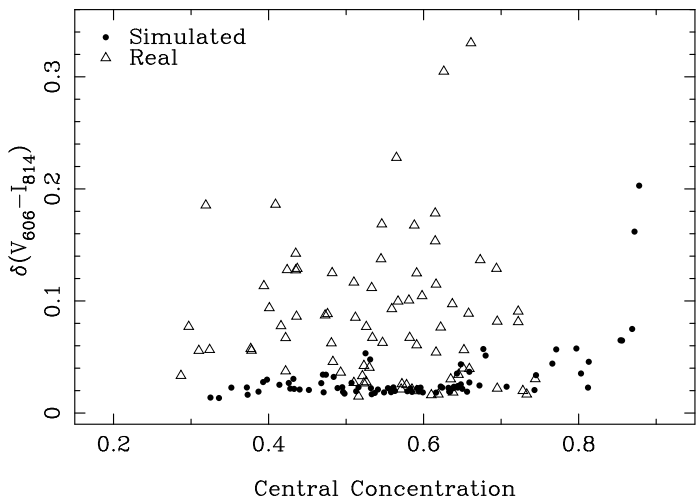}}
\end{center}
\caption {\em The colour inhomogeneity \dvi for real and simulated
data sets as a function of their central concentration indicating
the level of bias introduced by the wavelength-dependence of the
HST point spread function.}
\label{psf-simu}
\end{figure}

As the HST point-spread function (PSF) varies as a function of
wavelength, spurious centrally-concentrated inhomogeneities might
be expected for sharply-peaked profiles such as those encountered
in spheroidal galaxies. In order to test this, we performed extensive 
Monte Carlo simulations using the {\sc Iraf} package {\sc artdata}.  
Artificial galaxy images were created and subsequently analysed using 
procedures identical to those used to analyse our observed data. 

Artificial spheroidal galaxies were synthesised with de Vaucouleurs
profiles, using stars from the \mbox{HDF-N} as PSF templates.
Simulated images were generated in both \ib and \vb --bands.  We
explored a range of half-light radius ($r_e$) from $0.04''- 0.65''$
corresponding to $\sim 0.4 - 6$~kpc ($H_0=65$~km~s$^{-1}$~Mpc$^{-3}$,
$q_0=0.1$), and a range of magnitudes representative of our sample
($18<I_{814}<24$ mag). As the influence of the PSF on the central
colours is expected to be a strong function of the ``peakiness'' of
the central portion of the galaxy, we chose to probe the influence of
the PSF on both real and simulated galaxies as a function of the
central concentration ($C$) parameter defined in Abraham \etal (1994).

Figure~\ref{psf-simu} shows the result of this exercise. The PSF
variations affects the \dvi statistic only for the most
centrally concentrated (essentially stellar) objects. Over the
observed range of {\em C} for the real data, $0.3\simlt C \simlt 0.7$,
\dvi in the simulated data is well below the values recovered from the
real observations, and we conclude that this effect will not seriously
affect any of the following conclusions.

\subsection{Evolution in Field Spheroidals}

Figure~\ref{fig-mosaic} shows \vi colour maps for the 79 spheroidals
in our HDF sample sorted in ascending order according to \ib
magnitude. Redshift and \dvi values are also marked in each case;
redshifts in parenthesis refer to photometric estimates. The most
striking characteristic of these data is the large proportion with
significant internal colour variations. We emphasise that we have not
significantly varied the "colour stretch'' from panel to panel (as
indicated in the colour bar) to exaggerate this result. Adopting, for
simplicity, a colour variation of $\simeq 0.2$~mag as indicative of
the effect (since this broadly corresponds to a change in visual
colour according to our look-up table), it appears that $\sim 40\%$ of
the field spheroidals show a detectable degree of inhomogeneity.  In
most cases where inhomogeneities are seen, these appear to be in the
cores. Obviously, the visual inhomogeneities correlate closely with
\dvi.

In \S~\ref{clusters-section} we will calibrate these inhomogeneities
with reference to a corresponding sample of cluster spheroidals, but 
it is meanwhile interesting to consider whether these internal colour 
variations correlate with independent constraints. In doing this we
will make the assumption that the bluer colours seen are indicative of 
recent star formation.   

\begin{figure*}
\begin{center}
\leavevmode \centerline{\psfig{file=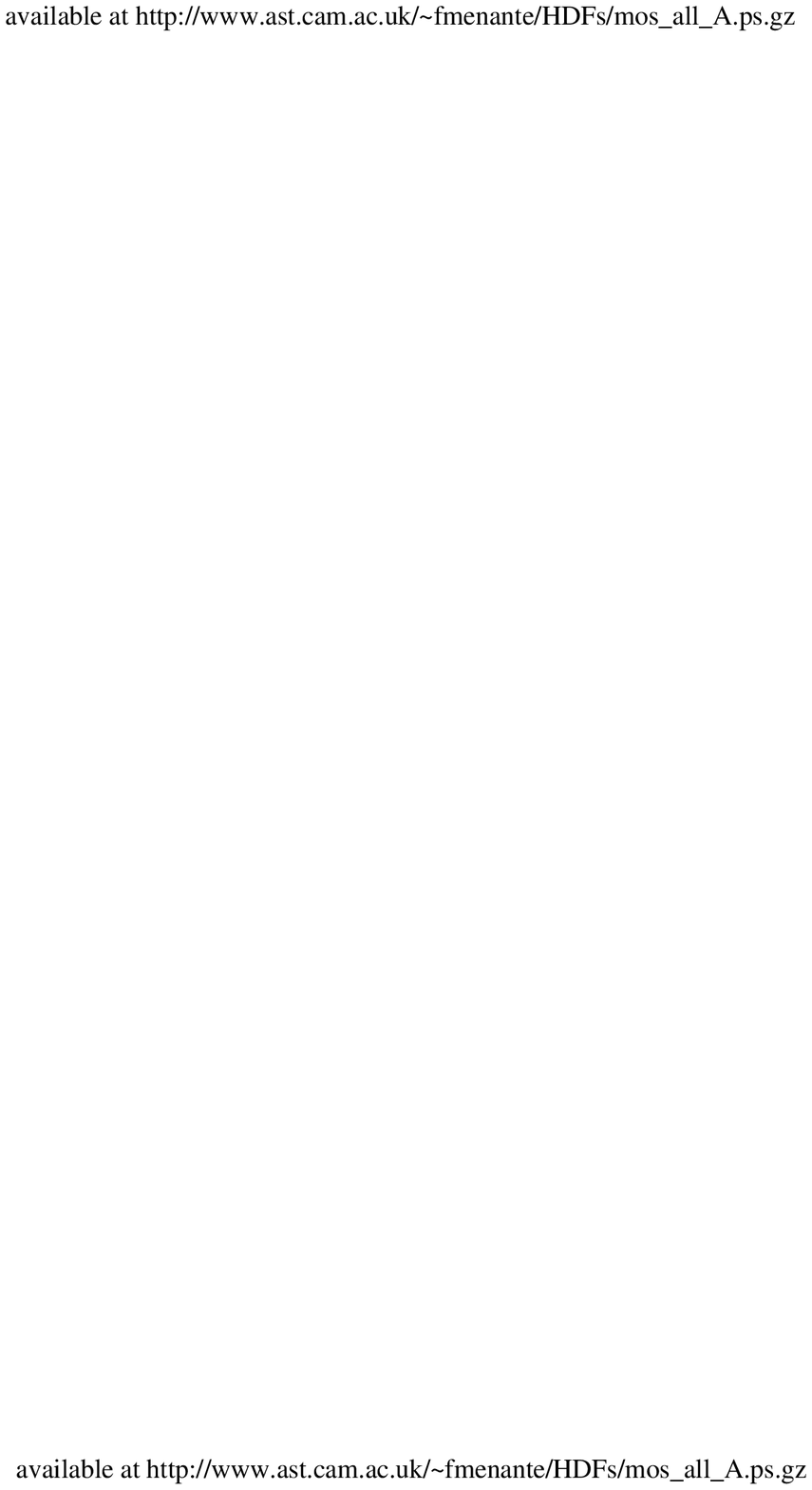,width=16cm,angle=0}}
\end{center}
\caption {\em Colour images of 79 HDF field spheroidals keyed to their \vi 
colours. The integrated \ib mag for each galaxy is shown in each sub-panel
together with the redshift and \dvi. Redshifts in parenthesis
indicate photometric estimates. Axes labels correspond to arcseconds}
\label{fig-mosaic}
\end{figure*}
% Image was too big, divided into 2 images.
\begin{figure*}
\begin{center}
\leavevmode
\centerline{\psfig{file=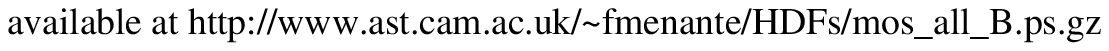,width=17cm,angle=0}}
\end{center}
\contcaption{\em }
\label{fig-mosaicB}
\end{figure*}

\subsubsection{Comparison with Kodama \etal}

\begin{figure}
\begin{center}
\leavevmode
\centerline{\psfig{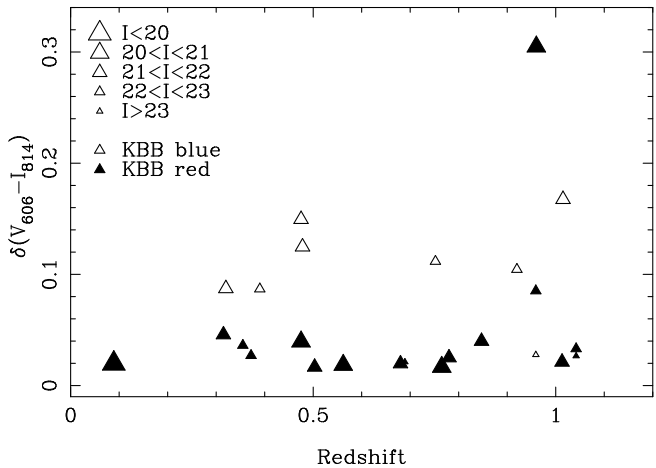}}
\end{center}
\caption {\em The distribution of \dvi with redshift for those HDF
spheroidals studied by Kodama \etal (1998, KBB). The size of the
symbols indicates the \iw integrated magnitude.}
\label{kbb-fig}
\end{figure}

We first compare our \dvi estimates with the results of Kodama, Bower
\& and Bell (1998, KBB). Those authors used stellar population models
and photometric redshifts to construct a rest-frame $U-V$ vs. $V$
colour-magnitude diagram for field spheroidals in the HDF-N. They took
the morphologically-classified sample of 35 spheroidals with $K<20.5$
from Franceschini \etal (1998) of which 29 are in our sample. The 6
missing cases correspond to two classes of objects: {\em a)} objects
that only Franceschini \etal classified as spheroidals, and {\em b)}
double detections in the Franceschini \etal (1998) study which we have
eliminated.  Using spectral synthesis models, KBB concluded that the
sample could be described as comprising a red sequence of old
passively-evolving systems, with an additional component (comprising 30\% of 
the total sample) with bluer optical-infrared colours that lay significantly
off the fiducial colour-magnitude relation.

We have divided the Franceschini \etal subsample of our \mbox{HDF-N}
spheroidals into 'red' and `blue' examples depending on whether their
rest-frame colours lie within the red sequence limits defined by KBB.
Specifically, we classed as blue all spheroidals whose rest-frame
colours have $U-V<0.8$. Figure~\ref{kbb-fig} compares our \dvi with the
red/blue classifications demonstrating a strong correlation with the
estimated $U-V$ rest-frame colour from KBB. In all but one case, red
KBB spheroidals correspond to low $\delta(V-I)$ systems in our scheme.
We estimate that $\sim 9$ out of of 29, or $\sim 30$\% of spheroidals
have values of \dvi suggesting some degree of star-formation activity,
a value in excellent agreement with the KBB estimate.

The strong correlation between our model-independent measures of \dvi
and KBB's estimates of rest-frame $U-V$ is reassuring evidence that
simple spectral synthesis models used to describe spheroidal evolution
offers a valid way to quantitatively interpret our results.  Armed with
this reassurance, in \S\ref{sfr-section} we will use similar models to
attempt to constrain the amount of star-formation contributed by
evolving spheroidals as a function of redshift.

\subsubsection{Low \dvi field spheroidals}
\label{lowdvi-section}

\begin{figure}
\begin{center}
\leavevmode
\centerline{\psfig{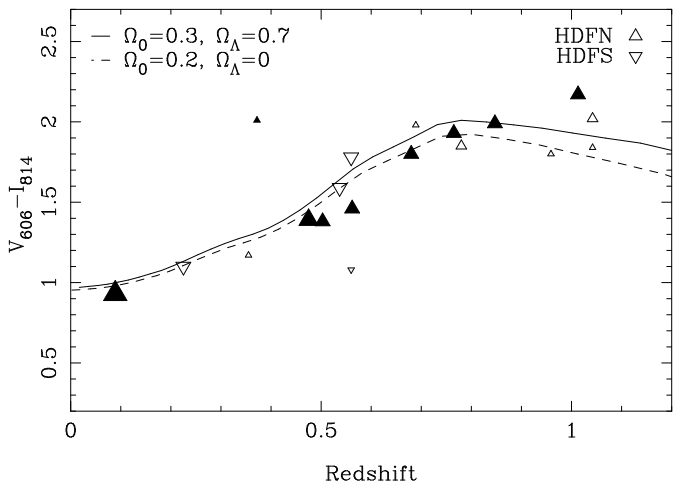}}
\end{center}
\caption {\em The \vi observed colours for a subsample of internally
homogeneous (\dvi$\leq0.04$) spheroidals as a function of redshift
compared with the predictions of a high-redshift single burst model
($z_F=5$, \mbox{$\tau=1$}~Gyr) and solar metallicity for a set of
cosmologies. Inverted triangles refer to HDF-S and unfilled symbols
refer to galaxies for which only photometric redshifts are available.}
\label{lowdvi-fig} 
\end{figure}

Despite the significant fraction of spheroidals showing some degree of
colour variation, the majority show homogeneous internal colours,
suggesting coeval evolution of their stellar populations. We selected
the galaxies with low \dvi$\leq0.04$ and compared their {\em
integrated} colours against the predictions of a monolithic collapse
model. As expected this sample closely tracks the expected
colour-redshift relation of a passively evolving
(Figure~\ref{lowdvi-fig}) providing reassurance that \dvi is a good
statistic from which to differentiate active and passive systems.

An interesting feature is the somewhat larger fraction
of homogeneous spheroidals within \mbox{HDF-N} with respect to the
HDF-S. Only 4 spheroidals in the HDF-S show low \dvi values
compared to 15 spheroidals in the HDF-N. This could be interpreted as 
evidence for clustering in the HDF-N as suggested by Kodama \etal (1998).  

\subsection{Evolution in Cluster Spheroidals}
\label{clusters-section}

As we discussed in $\S$1, it is now commonly-believed that spheroidals
in the cores of rich clusters formed the bulk of their stars before $z
\simgt 2$ (Bower \etal 1992, Ellis \etal 1997, Governato \etal 1998,
Kauffmann \etal 1999). Cluster samples therefore represent suitable
benchmarks whose \dvi values at a given redshift can be used to provide
a model-independent calibration of the internal colour scatter expected
from an old stellar population at a given redshift. 

\begin{figure}
\begin{center}
\leavevmode
\centerline{\psfig{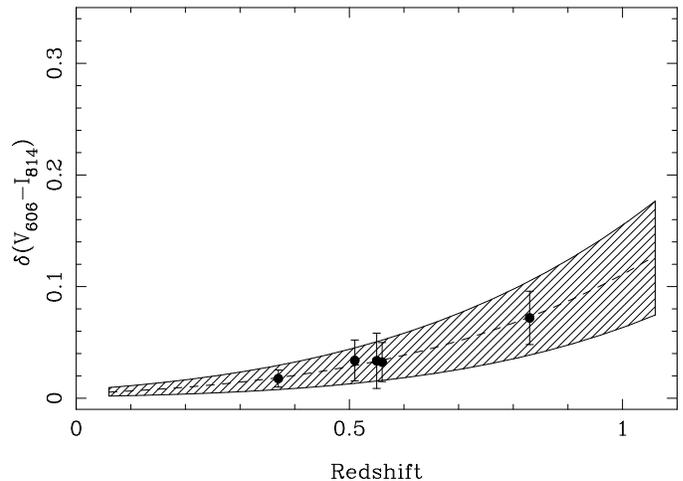}}
\end{center}
\caption {\em Mean \dvi values and $1\sigma$ dispersions calculated for 
spheroidal galaxies in 5 clusters observed with HST in similar
passbands to those used for the HDF field study.}
\label{field-cluster-fig}
\end{figure}

As mentioned earlier, our limiting magnitude in the cluster sample was
chosen as $I_{814} = 22$, considerably brighter than adopted for the
deeper HDF images. The morphological classifications were taken from
the published articles.  In the case of the MORPHS sample there is a
cross-check given one of us (RSE) classified both samples.  As in the
previous section we have calculated \dvi for each cluster spheroidal
using the prescription given in equation~\ref{dvi-eqn}. We find a
striking absence of internally inhomogeneous systems. In fact, we find
{\em no} cluster spheroidals with blue cores in {\em any} of the five
clusters.  We then calculate a median \vi value and will take this as
representative of the \dvi value appropriate for an old spheroidal
population over the redshift range $0.37<z<0.83$
(Figure~\ref{field-cluster-fig}).

\begin{figure}
\begin{center}
\leavevmode
\centerline{\psfig{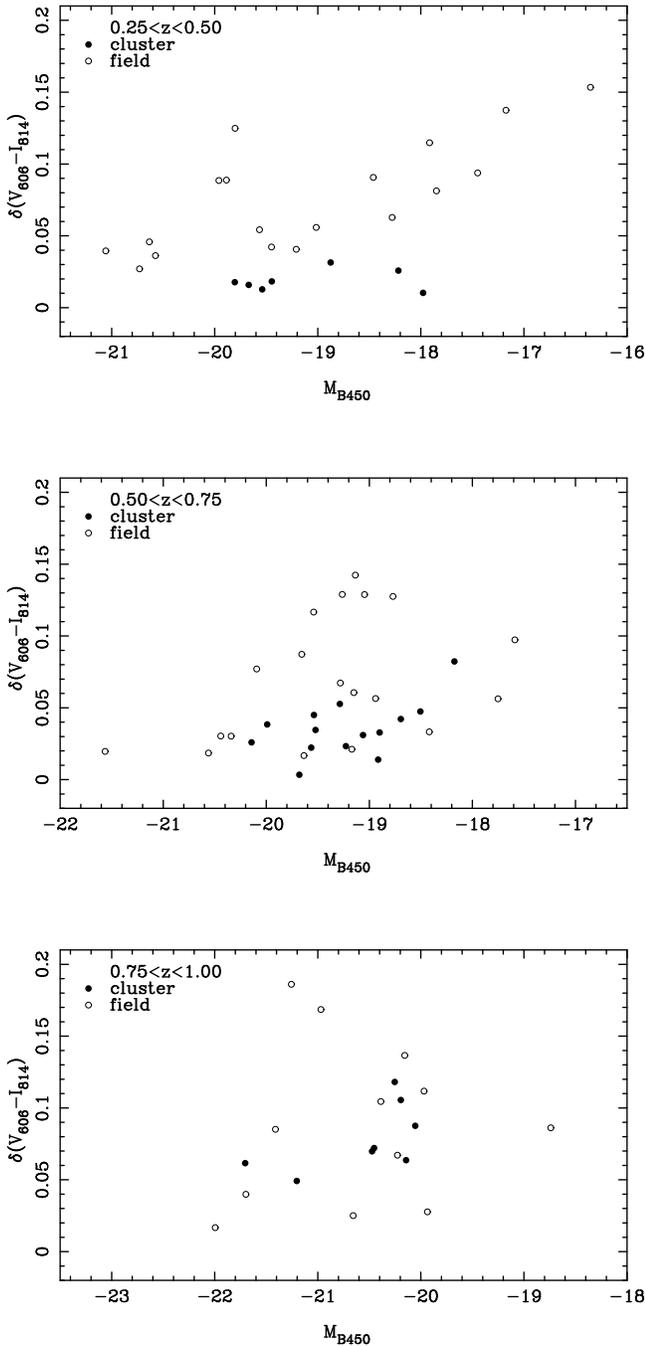}}
\end{center}
\caption {\em \dvi values for the \mbox{HDF-N} and \mbox{HDF-S} 
spheroidals compared with their clustered counterparts as a function of  
$M_{450}$ absolute magnitude for three redshift ranges. Solid points
represent cluster galaxies and open ones field spheroidals.}
\label{scatter-abs}
\end{figure}

The strongest result of this paper is that field spheroidals show
substantially greater internal colour dispersions than their clustered
counterparts. Since the limiting magnitude of the field sample is two
magnitudes brighter than that adopted for the cluster sample, this is
most cleanly demonstrated by comparing \dvi values on a
galaxy-by-galaxy basis in both samples as a function of rest-frame
luminosity (Figure~\ref{scatter-abs}). Absolute magnitudes were
calculated assuming \mbox{$H_0=65$~km~s$^{-1}$~Mpc$^{-1}$}. In the
cluster analysis, galaxies smaller than 60~pixels were not considered,
as the \dvi measurement was determined to be too noisy to be
meaningful. Although about 15\% of the cluster spheroidals were
rejected on this basis, from Figure~\ref{scatter-abs} we are reassured
that we are probing a similar range of $M_{B450}$ absolute magnitudes.
 
For the cluster spheroidals we calculated $M_{B450}$ for each galaxy
from its \ib apparent magnitude by applying {\em K}-corrections and colour
terms based on a high-redshift single-burst model ($z_F=5,
\tau=1$~Gyr) which, as we have discussed, reproduces the observed
colours (Bower \etal 1992, Ellis \etal 1997). For the field
spheroidals we computed $M_{B450}$ using \bb apparent magnitudes,
given in version 2 of the HDF {\tt SExtractor} catalogues, and a
maximum likelihood method describe more fully in
$\S$\ref{sfr-section}. Anticipating this discussion, briefly the
method allows us to obtain the most-likely \mbox{{\em K}-correction} for
each individual pixel within the galaxy. This complication is
necessary given the hybrid nature of the spectral energy distribution
in the internally inhomogeneous cases.

Clearly, over similar ranges of absolute magnitudes, field spheroidals
show markedly larger dispersions in their internal colours than do
cluster spheroidals. Although cluster spheroidals have more
homogeneous internal colours, there is a trend for \dvi to grow as a
function of redshift. The trends calculated in Figure 2 shows that
this is not an artifact relating to a greater sensitivity to very
small amounts of star formation. Unfortunately the limited depth of
the cluster data does not allow us to put more meaningful constraints
on the evolution of the cluster population.

\section{SPECTRAL SYNTHESIS MODELLING OF INTERNAL COLOUR VARIATIONS}
\label{model-sec}

Simple comparisons between the internal colour distributions of field
and cluster samples allow us to establish, in a model-independent way,
the proportion of active systems as a function of redshift and
environment.  Given that detailed spectroscopic information is not yet
available for the sample, in order to learn more about the relevant
timescales for the presumed star formation activity we must now turn
to more detailed synthesis models.  In this section we will
investigate how simple models can be used to place broad constraints
on the buildup of mass implied by the internal colour variations
reported in this paper.

\subsection{A Simple Evolutionary Model For Field Spheroidals}

\begin{figure}
\begin{center}
\leavevmode
\centerline{\psfig{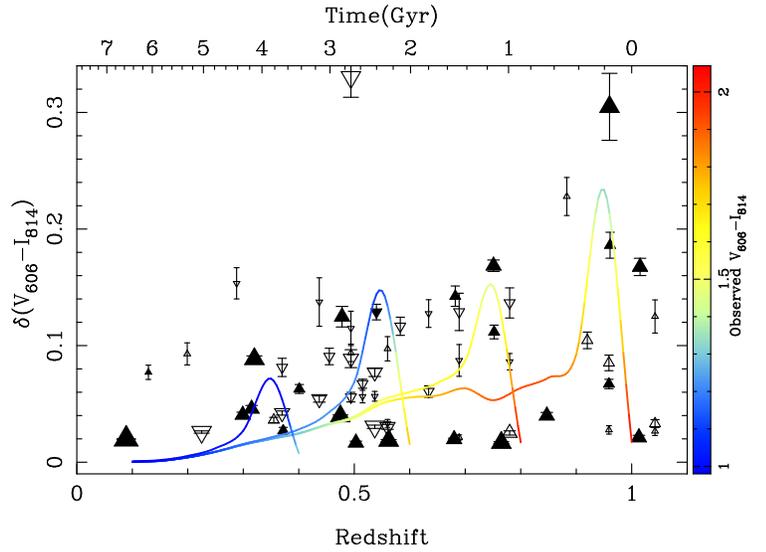}}
\end{center}
\caption{\em The observed \dvi predicted for the blue-nucleated model
spheroidal (solid line) compared with observed values in the HDFs. The
colour gradient along the evolutionary track maps the observed
evolution in \vi according to the right hand scale bar.  Error bars in
the field spheroidals show the $1\sigma$ deviation obtained from 100
bootstrap resamplings of the data pixels within each galaxy and field
symbols follow the designations of Figure 6.}
\label{burst-fig}
\end{figure}

The blue light in field spheroidals reported in the present paper
arises, in the majority of cases, in the central regions. Possible
origins include residual star-formation from a recent major merger
(Kauffmann 1996), low-level star-formation that is a relic from a
high-redshift collapse (Jim\'enez \etal 1999), cooling flows from hot
gas surrounding the central region of the ellipticals, or a bursts of
central star-formation possibly related to the inflow of gas arising
from a recent minor merger. In this section we aim to place
constraints on the star-formation timescales and masses associated
with the activity we claim to observe predominantly in the field
population.  We do this within the context of a simple model that
seeks to reproduce the observed spatially resolved colours of the
subset of active ellipticals with prominent blue nuclei.

We assume that an intermediate redshift field spheroidal can be
described with a two component model -- an old underlying stellar
component which formed at high redshift and a secondary younger
component (responsible for the central blue light). The integrated
colours of both components are estimated using the stellar population
synthesis library of Bruzual \& Charlot (1996, BC96). We modelled the
old component using a 1 Gyr exponentially declining burst of star
formation commencing at a redshift of formation $z_F=5$. We adopt a
flat Universe with \mbox{$\Omega_0=0.3$},
\mbox{$\Omega_{\Lambda}=0.7$}. The secondary population is modelled
via an additional instantaneous burst of star-formation presumed to
occur close to the epoch of observation. The present modelling is
similar to the one described in $\S$~\ref{estimator}, however here we
include the evolution as function of the galaxy age.

In order to mimic the resolved properties of a model galaxy, we create
an image of the two galaxy components using a modified version of {\sc
IRAF mkobjects}. For the old component we use a de Vaucouleurs profile
with a half-light radius of \mbox{$r_e=2.5$~kpc}. For the central blue
component we also use a de Vaucouleurs profile, but with an effective
radius $0.1r_e$. We assume 15\% of the galaxy mass is in the young
component, and that the total mass is $M=1\times
10^{12}$~M$_{\sun}$. Finally, between $0<z<z_{burst}$ we combine the
two components by adding up the images using the colours predicted by
our spectral synthesis models.  For the resulting galaxy we calculate
$\delta(V-I)$ and repeat the procedure four times corresponding to
15\% (by mass) bursts at redshifts $z_{burst} =$ 1.0, 0.8, 0.6 and
0.4.

Figure~\ref{burst-fig} shows the result of this exercise, superposed
on the observed \dvi for the HDF spheroidals. Error bars correspond to
$1\sigma$ estimates obtained by bootstrapping the data as described in
Efron \& Tibshirani (1986). We conclude that our simple model can
successfully reproduce the range of observed \dvi as well as the
observed increase \dvi with redshift. The essential point is that the
observed internal colour scatter is not necessarily a relic of the
initial elliptical formation event and can quite easily represent a
perturbation introduced by a recent episode of star-formation (forming
only a modest proportion of the total stellar mass) superposed on a
pre-existing old population. If this is the case then excursions from
the homogeneous colours of the underlying old stellar population are
probably brief, lasting about 1 Gyr after which the galaxy returns to
its pre-burst value. Ultimately high resolution spectra of large
sample of spheroidals using Balmer lines will enable us to disentangle
the {\em internal clock} of star formation for field spheroidals.
However from this simple study it seems reasonable to conclude that as
many as half the field spheroidal population share this activity by
$z\sim1$.

\subsection{The Star Formation History of Spheroidals}
\label{sfr-section}

If our interpretation of Figure~\ref{burst-fig} is correct, the
question arises as to the extent of this recent activity in terms of
the evolution of global star formation and its role for the mass
assembly of field spheroidals as predicted in popular hierarchical
models. Although the HDF samples are very modest ones on which to base
such fundamental studies, we explore these wider issues using the
analytical techniques described in Abraham \etal (1999).

In the \mbox{HDF-N}, Abraham \etal were able to reconstruct a simple
cosmic star-formation history by interpreting resolved multi-colour
data in the context of evolutionary synthesis models. For each HDF
galaxy, an integrated star formation rate was determined and
volume-averaged values (\sfrdensity) for the entire population was
derived using a simple $V/V_{max}$ methodology. Here we will attempt
to estimate the contribution to \sfrdensity arising from our HDF
spheroidals and compare its evolutionary decline with time with that
observed in other morphological populations. The crucial advantage of
an analysis utilising the resolved colours is that it enables a
determination of the relative ages and star-formations of the
components {\em within} the galaxies.

\subsubsection{The Method}
\label{sfr-method-sec}

\begin{figure*}
\begin{center}
\leavevmode \centerline{\psfig{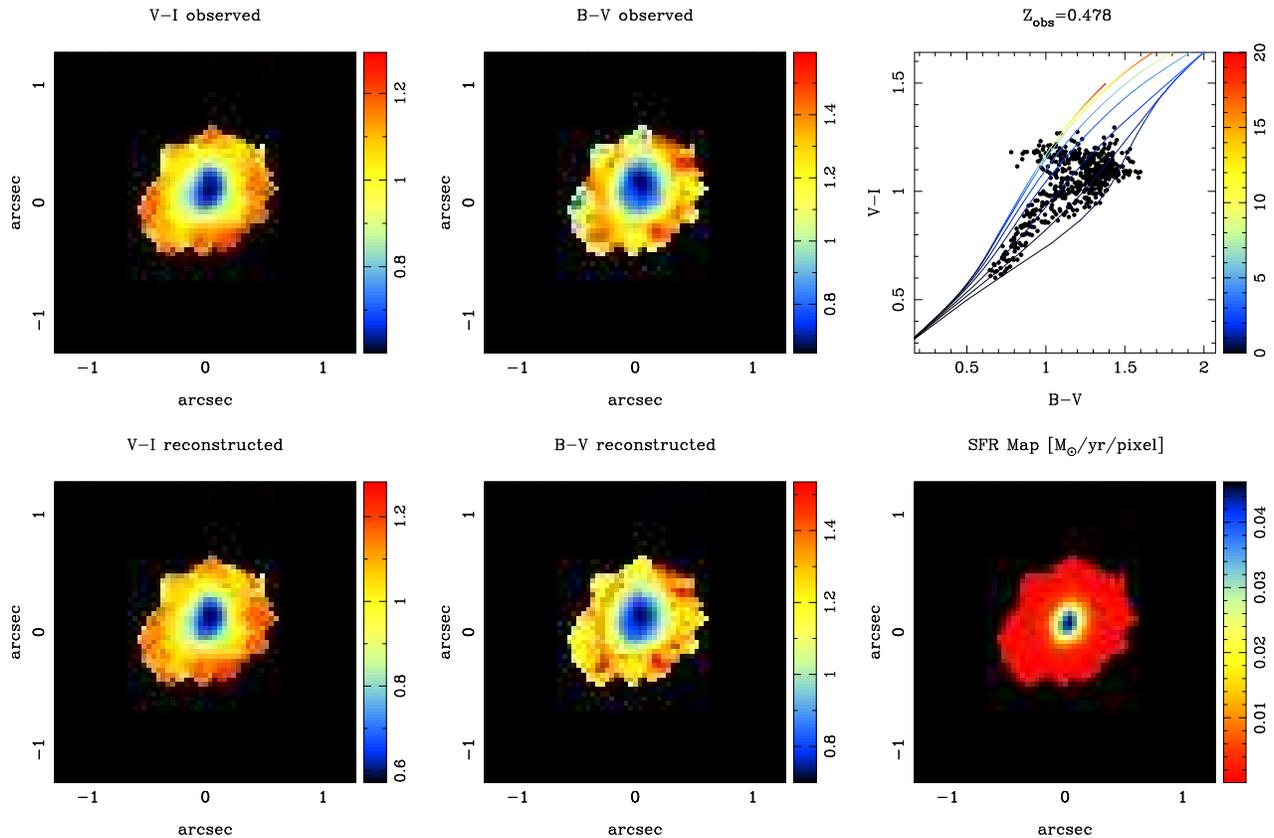}}
\end{center}
\caption {\em A demonstration of our maximum likelihood methodology
showing the observed and reconstructed \vi and \bv colours. The
reconstructed colours and SFR (lower right) maps were based on the the
best fit models to the pixel colour-colour diagram (upper right). The
colour assigned to the tracks drawn in the colour-colour diagram
indicates age in Gyr according to the right hand colour bar
(blue=young, red=old). As expected, the blue core is a region of
current star formation, whereas the outer region shows virtually no
activity.}
\label{sfr-example}
\end{figure*}

Using the precepts of Bruzual \& Charlot (1991), the total flux from a
stellar system, $F_{\lambda}$ observed at an age $t_0$ can be
described as the convolution of the spectrum of an evolving
instantaneous burst, $f_{\lambda}(t)$, with an star-formation history
given by $\Psi(t)$.
\begin{equation}
F_{\lambda}(t_0) = \int_{0}^{t_0} \Psi(t_0-t)f_{\lambda}(t) dt
\label{eqn-sfr}
\end{equation}
In our analysis, we use the stellar population synthesis libraries
from {\sc gissel96} (Bruzual \& Charlot 1996) to estimate
$f_{\lambda}$, assuming a Scalo IMF with lower and upper limits of 0.1
and 125~M$_{\sun}$ respectively and solar metallicity. For the SFR
function $\Psi(t)$, we assume that the colours of a given stellar
population can be modeled by exponential star-formation histories with
a characteristic time-scale $\tau$ of the form,
\begin{equation}
\Psi(t) = \Psi_0e^{-t/\tau}.
\end{equation}
which enables us to describe a wide range of star-formations, given
that $\tau \to \infty$ approximates a constant star formation, while
$\tau \to 0$ approximates an instantaneous burst.

As the $U_{300}$-band flux is weak for our sample, adding this to the
analysis would enlarge the uncertainties in the observed colours of
spheroidals rather than further constrain our estimates. Therefore, we
concentrate on the $B_{450}$, $V_{606}$, and $I_{814}$ bands only. We
calculate the \bv vs \vi evolutionary tracks as observed at the galaxy
redshift for a set of representatives formation histories ($\tau =
0.1$, 0.2, 0.4, 1, 2, 4, and 9~Gyr) using the stellar population
libraries from {\sc gissel96}. For each spheroidal, we select all the
contiguous pixels within a surface brightness threshold of
$B_{450}=26$~mag/arcsec$^2$, and from the selected pixels construct
the corresponding \bv vs \vi colour-colour diagram for each of the
galaxies. We define this limit in order to maximize the signal/noise
associated per pixel in both the \bv and \vi colours, selecting in the
lowest signal band in order to ensure an uniform signal criteria in
the pixel selection. We estimate the optimum model and age for each
pixel using a maximum likelihood estimator ${\cal L}$ defined as:
\begin{equation}
{\cal L}(t,\tau) = \prod_{i=1}^{2} {1\over \sqrt{2\pi}\delta C_i}
 \exp\left( - {(c_i-C_i)^2\over 2 \delta C_i\,^2}\right)
\end{equation}
where we take the product over both colours \bv and \vi, $C_i$ and
$\delta C_i$ represent the data colour and colour error associated
with the pixel respectively, and $c_i$ is the colour of the modelled
track at a given age and {\em e}-folding time-scale $\tau$. For each
pixel we obtain the most likely age, SFR and $\tau$ value. Repeating
over all acceptable pixels, we obtain the integrated value at the
observed redshift. This procedure is repeated for all spheroidals in
the HDFs.  In order to illustrate this methodology, in
Figure~\ref{sfr-example} we show a typical pixel-by-pixel
colour-colour diagram and illustrate how the predicted evolutionary
tracks at the redshift of observation can reconstruct the \vi and \bv
observed colours and construct a SFR map for each galaxy pixel.

Given that our methodology depends explicitly on the redshift of the
galaxy, we need to take into account the uncertainties associated with
the photometric estimates when calculating SFR for galaxies with
photometric redshifts. Up to $z\sim1$, photometric redshifts have
proved to be in reasonable good agreement with spectroscopic redshifts
with an rms error of $\delta z\sim0.1$ for red populations (Hogg \etal
1998, Gwyn \etal 2000, de Soto \etal 1997). We performed Monte Carlo
simulations by refitting each galaxy using the models for the
appropriate photometric redshift plus a random component of $\delta
z=0.1$. Repeating this exercise for each galaxy with photometric
estimates allows us to obtain an uncertainty on the total SFR
estimated for those systems for which only photometric redshifts are
available.

Clearly the predictions of spectral synthesis models are sensitive to
both metallicity and dust content, neither of which is well
constrained in our data. The presence of dust would have the effect of
underestimating the computed SFR, whereas metallicity changes will
produce systematic shifts between the predicted age and colours of
each galaxy. Whilst recognising the limitations involved, our primary
goal at this stage is to illustrate the potential of our resolved
colour technique to determine redshift-dependent trends rather than to
derive precise absolute values. Throughout the SFR analysis we will
use solar metallicity in our calculations, as from
Figure~\ref{lowdvi-fig} we can see that this reproduces fairly well
the observed colour of low \dvi spheroidals. With resolved infrared
data such as may be available from NICMOS, it may ultimately be
possible to limit the uncertainties arising from dust.
  
\subsubsection{Evolution in the Star-Formation density of spheroidals}
\label{sfr-density-sec}

We now estimate the SFR density ($\rho_{\mathrm SFR}$) of spheroidals as a 
function of redshift and choose to compare these estimates with similar
ones derived for a parallel sample of morphologically-irregular
galaxies in both HDFs. In the case of the latter, we used the
morphological classification of Abraham \etal and adopted the
same \ib limit for both HDFs.

\begin{figure}
\begin{center}
\leavevmode
\centerline{\psfig{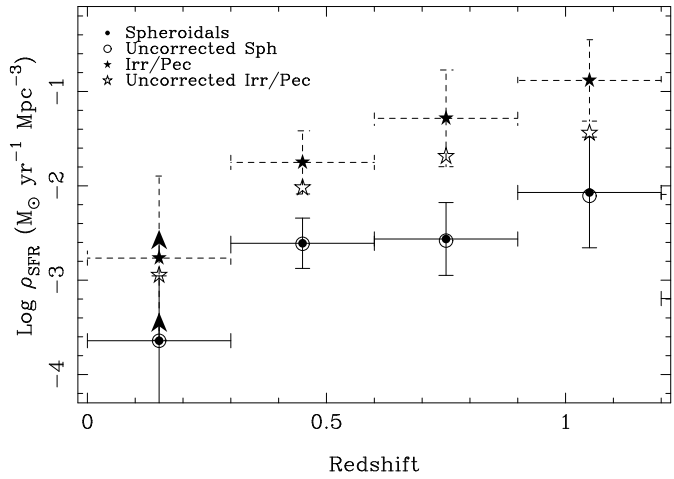}}
\end{center}
\caption{\em The comoving SFR density computed using the 63 spheroidal
(filled circles) and 51 morphologically-irregular (filled stars)
galaxies in both HDFs. Corrections for incompleteness were made
assuming SFR $\propto L$, local luminosity functions and evolutionary
corrections described in the text.  Hollow symbols represent
uncorrected values. Error bars take into account systematic
uncertainties implied in the calculation of the SFR in individual
galaxies and bootstrap errors which account for the small sample
size.}
\label{sfr-density}
\end{figure}

We calculate the SFR density at a given redshift interval $z_1<z<z_2$
using the expression,
\begin{equation}
\rho_{\mathrm SFR}(z) = \sum_{i; z_1<z_i<z_2} \frac{{\mathit SFR}_i}{V_{max,i}}
\end{equation}
where the sum is over all the galaxies $i$ for which $z_1<z_i<z_2$ and
$V_{max}$ is the volume at $z_{max}$, the largest redshift at which
the galaxy $i$ remains brighter than our limiting magnitude (Schmidt
1968). Clearly as we move towards higher redshifts our sample becomes
increasingly biased towards intrinsically more luminous galaxies and
we will consequently neglect a contribution to \sfrdensity from
sub-luminous objects. To correct for this we need to assume: {\em i)}
the dependence of the SFR on the luminosity $L$, {\em ii)} the {\em
shape} of the spheroidal luminosity function and {\em iii)} the
luminosity evolution assumed for the spheroidal population. Whilst
none of these is unfortunately known precisely, we can make progress
by assuming that $\mathit{SFR}\propto~L$ (i.e. the most luminous
galaxies form the most stars) and adopting a suitable Schechter
luminosity function (LF). Under these assumptions, the corrected SFR
density ($\tilde{\rho}_{\mathrm SFR}$) would become:

\begin{equation}
\tilde{\rho}_{\mathrm SFR}(z) = C(z)\rho_{\mathrm SFR}(z)
\end{equation}
with,
\begin{equation}
C(z) = \frac{\int_0^{\infty} {\mathit{SFR(L)}} \phi(L) dL}
            {\int_{L_{min}}^{\infty} {\mathit{SFR(L)}} \phi(L) dL}
     = \frac{\Gamma(\alpha+2)}{A}
\label{eq-correction}
\end{equation}
and,
\begin{equation}
A = 0.4\ln(10)\int_{-\infty}^{M^-} 10^{0.4(M^{\star}-M)(\alpha+2)} e^{-10^{0.4(M^{\star}-M)}}dM
\end{equation}
where $M^\star$ may be permitted to evolve with redshift (e.g.
according to an evolutionary correction) and $M^-$ and $L_{min}$ are
the faintest absolute magnitudes and luminosities respectively within
the redshift interval in question.
 
As an illustration of the technique, we adopted Schechter LFs from
Marzke \etal (1998) with \mbox{$M^\star_{b_j}=-20.87$} and
\mbox{$\alpha=-1$} for the HDF \mbox{spheroidals} and
\mbox{$M^\star_{b_j}=-21.28$} and \mbox{$\alpha=-1.81$} for the
irregulars. For the spheroidals we assumed $M^*$ evolves according to
that appropriate for a passively evolving system with
\mbox{($\tau=1$~Gyr)}. In the case of the irregulars, we assumed a
constant SFR \mbox{($\tau=5$~Gyr)}.

Finally is necessary to define a number density correction factor,
${\cal K}(z)$, very similar to the one described in
equation~\ref{eq-correction} to estimate the redshift completeness of
our sample, viz.
\begin{equation}
{\cal K}(z) = \frac{\Gamma(\alpha+1)}{B} 
\label{eq-correction-number}
\end{equation}
and
\begin{equation}
B = 0.4\ln(10)\int_{-\infty}^{M^-} 10^{0.4(M^{\star}-M)(\alpha+1)} e^{-10^{0.4(M^{\star}-M)}}dM
\end{equation}
where $M^-$ is the faintest absolute magnitude included in the sample
at a given redshift determined by our flux limit of $I_{814}=24$~mag
and the distance modulus relation. 

Although ${\cal K}(z)$ and $C(z)$clearly depend on the
poorly-constrained LF shape, for our adopted values we find
completenesses (in terms of the contribution of detectable galaxies to
the integrated \sfrdensity) of $\sim 90\%$ and $\sim 80\%$ for 63 HDF
spheroidal and 51 HDF morphologically-irregular galaxies respectively.

Figure~\ref{sfr-density} shows the results of our calculations and the
uncertainties introduced by our corrections for incompleteness.  As
expected these corrections become more important at higher redshifts
and for the morphologically-irregular population whose LF is presumed
to be steep at all times ($\alpha<-1$). However in most of the cases
the corrections are comparable to the size of the statistical
uncertainties. Random errors bars were calculated by adding in
quadrature the systematic errors in the SFR calculation for individual
galaxies, including the Monte Carlo estimates for the photometric
redshift and Poisson errors estimated by bootstrapping 100 times the
sample when computing the SFR density.  In Table~\ref{sfr-table} we
list the computed values for \sfrdensity as well as their
uncertainties.

Figure~\ref{sfr-density} enables us to provide the first rough
constraints on the contribution of the increased colour variation in
the HDF spheroidals to the evolution of the $\delta(V-I)$. A rise with
redshift is discernible but the uncertainties are
considerable. Interestingly, the evolution in \sfrdensity mirrors, but
at a lower level, that established more convincingly for field
irregulars from the CFRS/LDSS sample of Brinchmann \etal (1998) (but
estimated here for the HDF from a completely independent resolved
colour method). Brinchmann \etal computed \sfrdensity between
$0.3<z<0.9$ using the equivalent width of {\sc[Oii]}. Although their
\sfrdensity values are systematically lower for both the spheroidal
and peculiar/irregular population, similar evolutionary trends were
observed in the overlapping redshift range $0.4<z<1.0$. If it could be
established that the decline in the spheroidal \sfrdensity was driven
by that associated with the demise of the more active irregular
population, this would provide further support that the blue cores
observed in the HDF spheroidals arise from recent mergers of the
irregular population (Le F\`evre \etal 1999, Brinchmann 1999, Brinchmann
\& Ellis 2000).
 
\begin{table*}
\begin{minipage}{140mm}
\footnotetext[1]{Single Burst ($\tau=1$~Gyr), $z_F=5$, Marzke \etal E/S0 LF ($\alpha=-1$, $M^*=-20.87$)}
\footnotetext[2]{Constant SFR ($\tau=5$~Gyr), Marzke \etal Irr/Pec LF ($\alpha=-1.81$, $M^*=-21.28$)}
\caption{SFR density estimates}
\label{sfr-table}
\begin{tabular}{@{}ccccc}
\hline
Redshift & Spheroidals$_{uncorrected}$ & Spheroidals$^{a}_{corrected}$ & Irr/Pec$_{uncorrected}$ & Irr/Pec$^b_{corrected}$\\
\hline\hline
 0.15 & $-3.642^{-2.956}_{-4.329}$ & $-3.641^{-2.954}_{-4.328}$ & $-2.948^{-1.001}_{-4.895}$ & $-2.766^{-1.896}_{-3.636}$\\
 0.45 & $-2.615^{-2.348}_{-2.881}$ & $-2.609^{-2.342}_{-2.875}$ & $-2.020^{-1.685}_{-2.354}$ & $-1.752^{-1.417}_{-2.087}$\\
 0.75 & $-2.581^{-2.195}_{-2.966}$ & $-2.563^{-2.177}_{-2.949}$ & $-1.688^{-1.176}_{-2.201}$ & $-1.285^{-0.772}_{-1.797}$\\
 1.05 & $-2.106^{-1.520}_{-2.692}$ & $-2.070^{-1.484}_{-2.656}$ & $-1.442^{-1.010}_{-1.873}$ & $-0.883^{-0.452}_{-1.314}$\\
\hline
\end{tabular}
\end{minipage}
\end{table*}

\section{CONCLUSIONS}
\label{conclusions}

We have exploited HST's unique capabilities to resolve distant galaxies by
examining the internal colour distribution of a sample of 79 field spheroidals
in both Hubble Deep Fields. From a model-independent analysis of our
data, we find:

\bigskip

$\bullet$ A significant fraction ($>$30\%) of field spheroidals show
marked variations in their internal colours. In most of the cases where
colour variations are seen, this manifests itself in terms of centrally-located
blue cores.

$\bullet$ Using a statistic, $\delta(V-I)$, to characterise the
internal homogeneity of a galaxy we have extended our analysis to
permit a comparison of the internal variations found in the HDF sample
with that for a sample of spheroidal galaxies in five rich clusters
spanning a similar redshift range. Marked differences are found in the
sense that galaxies of the same luminosity and redshift are
consistently more uniform in rich clusters than in the field.  Assuming
the blue colours located in the HDF field ellipticals arise from recent
star formation, our comparison provides strong evidence for more
extended star formation histories in field ellipticals (as expected in
hierarchical models).

$\bullet$ We compare our observational diagnostics with those used by
other workers and find general agreement. The most straightforward
interpretation of our data is that a fraction of field spheroidals at
intermediate redshift has undergone recent star formation, perhaps as
a result of recent merging or an inflow of material.

\smallskip

Using spectral synthesis models and a maximum likelihood analysis
developed by Abraham \etal (1999) we attempt to reproduced the
observed colour variations for the field spheroidals and to compute
their current SFR.  From this model-dependent analysis we find that:

\smallskip

$\bullet$ Assuming that the observed blue nuclei observed in some
field spheroidals arises from a recent burst of star-formation superposed
on a pre-existing old stellar population, we can readily reproduce the 
observed distribution of \dvi values. The star formation
episodes implied are short, lasting less that 1~Gyr, and typically involve
15\% of the galactic mass.

$\bullet$ If the HDF fields are representative, the fraction of field
ellipticals undergoing activity of this kind could be as high as
$\sim50\%$ at $z\sim1$.

$\bullet$ The contribution to the global SFR density from field spheroidals, 
whilst uncertain to determine accurately because of the necessary corrections 
for incompleteness and possible luminosity evolution, appears to 
rises modestly as a function of redshift in a manner which may connect
with the demise in activity associated with morphologically-irregular
systems. 

With a larger sample, augmented with diagnostic spectroscopy capable
of constraining the {\em timescale} of recent activity, it may be
possible to strengthen the connection between the declining luminosity
density contributed by morphologically-irregular galaxies and the
implied continued assembly of field ellipticals. The goal remains an
important one in understanding the origin of the Hubble sequence and
in defining the role of the environment (Brinchmann \& Ellis 2000).

\section*{ACKNOWLEDGMENTS}

We thank Jarle Brinchmann, Piero Madau and Neil Trentham for useful
discussions and suggestions. FM acknowledges support from an Isaac
Newton Scholarship and would like to thank Fundaci\'on Andes for
financial support.

\bsp % ``This paper has been produced using the ...''
\label{lastpage}

\end{document}